# Direct Estimation of Pediatric Heart Rate Variability from BOLD-fMRI: A Machine Learning Approach Using Dynamic Connectivity[*]

Abdoljalil Addeh [1-3,5], Karen Ardila [1-3,5], Rebecca J Williams[6]
G. Bruce Pike,[3-5] M. Ethan MacDonald [1-3,5]

[1] Department of Biomedical Engineering, Schulich School of Engineering, University of Calgary, Canada
[2] Department of Electrical & Software Engineering, Schulich School of Engineering, University of Calgary, Canada
[3] Department of Radiology, Cumming School of Medicine, University of Calgary, Canada
[4] Department of Clinical Neurosciences, Cumming School of Medicine, University of Calgary, Canada
[5] Hotchkiss Brain Institute, Cumming School of Medicine, University of Calgary, Canada
[6] Brain-Behaviour Research Group, University of New England, Australia

**Synopsis:**

**Motivation:** In many pediatric fMRI studies, cardiac signals are missing or have poor quality. Therefore, it would be highly beneficial to have a tool to extract Heart Rate Variation (HRV) waveforms directly from fMRI data without the need for peripheral recording devices.

**Goal:** Develop a machine learning framework to accurately reconstruct HRV tailored for pediatric applications.

**Approach:** A hybrid model using one-dimensional Convolutional Neural Networks (1D-CNN) and Gated Recurrent Units (GRU) analyzed BOLD signals from 628 ROIs, integrating past and future data.

**Results:** Achieved an 8% improvement in HRV accuracy, evidenced by enhanced performance metrics, demonstrating the model's effectiveness.

**Impact:** This method enhances pediatric fMRI by eliminating the need for peripheral photoplethysmography devices, reducing costs and simplifying procedures. It could also improve the robustness of pediatric fMRI studies, which are more affected by physiological and developmental variations than in adults.

## Introduction

Recent advancements in machine learning (ML) have facilitated a novel approach that leverages resting-state BOLD-fMRI signals to directly estimate respiratory variation (RV) waveforms in pediatric participants [1], thereby circumventing the need for external measurements. This methodology, validated with data from the Human Connectome Project in Development (HCP-D) [2], has proven highly effective in reconstructing RV. In contrast, the estimation of heart rate variation (HRV)—vital for addressing cardiac confounds—has not yet been explored in pediatric fMRI research. The challenge of HRV estimation is accentuated by its complex temporal dynamics, where BOLD-fMRI signal alterations span a substantial temporal window of 6 to 42 seconds and affect both gray and white matter [3-5].

In response to these challenges, this work presents the first study aimed at estimating HRV from resting-state BOLD-fMRI data in pediatric subjects. We propose an ML framework utilizing dynamic functional connectivity (dFC)-based brain atlases [6] designed to enhance the model's ability to capture the intricate temporal dynamics of HRV by integrating data from both gray and white matter. This approach is expected to refine the accuracy of HRV estimation, with an anticipated improvement in performance metrics by at least 5%, compared to static functional connectivity-based brain atlases used in the RV estimation study.

**Method**

This study employed a novel HRV reconstruction approach using a hybrid model of one-dimensional Convolutional Neural Networks (1D-CNN) and Gated Recurrent Units (GRU) to analyze BOLD signals from 628 ROIs (518 cortical, 62 subcortical, and 48 white matter ROIs) derived from dynamic and diffusion tensor imaging-based atlases [6, 7]. A 65 TR sliding window technique is used to estimate HRV at the 10th point, enhancing the capture of HRV-related BOLD fluctuations (Figure 1). Out of 2451 scans in the HCP-D dataset, 352 were selected for high-quality photoplethysmography (PPG) signals.

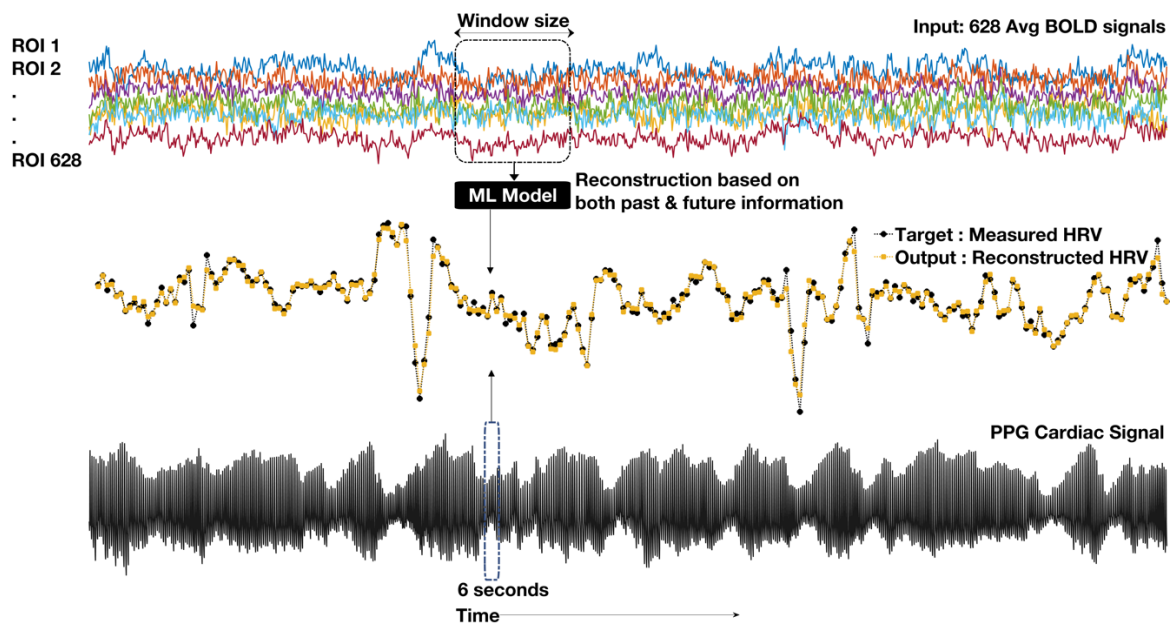

**Figure 1.** HRV Reconstruction Method from BOLD Signals: This schematic demonstrates our approach using averaged BOLD time series from 628 ROIs. HRV is estimated from a 65 TR sliding window at the 10th point, utilizing past and future data to capture the inherent dynamic fluctuations, typically around 40 seconds. The input matrix is formatted as [65 × 628], representing the comprehensive involvement of gray matter (518 ROIs), subcortical (62 ROIs), and white matter (48 ROIs) regions.

We used ten-fold cross-validation, dividing the dataset into ten subsets. Nine subsets were used for training and one for testing in each fold, repeated across all folds to create 10 independent models, with each subset tested once. Model performance was evaluated using multiple metrics: Mean Absolute Error (MAE), Mean Squared Error (MSE), Pearson correlation coefficient, and Dynamic Time Warping (DTW) on the test datasets. Performance metrics from all iterations were averaged, and the results were reported as mean values to provide a clear indication of the model's overall efficacy.

**Results**

Many PPG signals from the HCP-D dataset exhibit spikes from subject movement or technical issues. Figure 2 displays PPG signals, with the first (Figure 2.a) showing correctable noise used to train and test our ML model, while others are discarded due to excessive noise or incomplete data. The 1D-CNN + GRU network's accuracy in reconstructing HRV time-series is shown in Figure 3, demonstrating high precision, especially during significant HRV fluctuations.

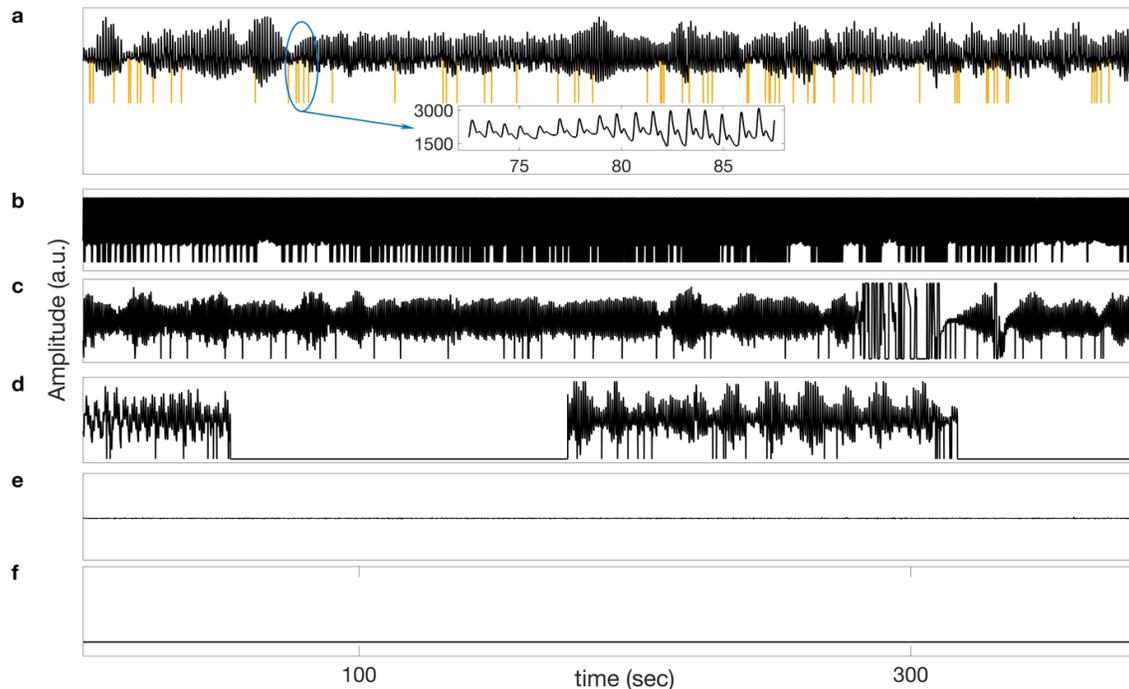

**Figure 2.** PPG signal variability in the HCP-D Dataset: This figure displays typical PPG signal issues found in the HCP-D dataset, common in pediatric fMRI studies due to subject movement or technical challenges. Panel (a) shows a signal with correctable spikes. Panels (b)-(f) show discarded signals: (b) uncorrectable spikes, (c) signal clipping, (d) partial recording with gaps likely due to sensor disconnection, (e) very small amplitude, unlike a normal PPG signal, and (f) no recording. Only group 'a' signals were used for machine learning, as they had adequate data quality.

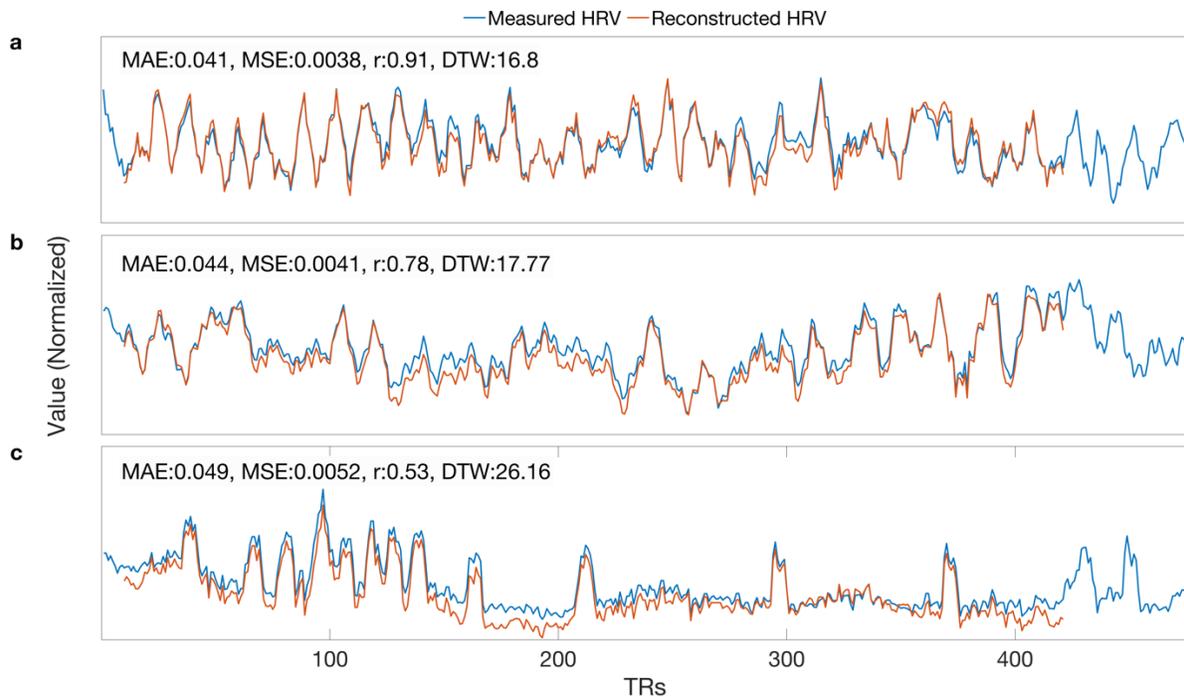

**Figure 3.** Evaluation of HRV reconstruction accuracy: This figure demonstrates the accuracy of HRV reconstruction using a hybrid 1D-CNN and GRU model across 628 ROIs. It compares measured and model-reconstructed HRV waveforms in three test cases (a, b, c), with corresponding MAE, MSE, Pearson correlation (r), and DTW values highlighted. These graphs highlight the model's ability to precisely capture HRV dynamics, especially during significant fluctuations, suggesting enhanced informational content in the BOLD signals about HRV.

Figure 4 details the relationship between HRV time-series standard deviation and the correlation of actual versus reconstructed HRV. Figure 5 evaluates the model's performance across different configurations:

- Dynamic Functional + White Matter ROIs: 580 dynamic and 48 white matter ROIs
- Dynamic Functional ROIs Only: 580 ROIs
- Static Functional + White Matter ROIs: 360 static [8] and 48 white matter ROIs
- Structural ROIs: 69 ROIs [9]

Configurations integrating dynamic and white matter ROIs outperformed others, confirming the benefits of this approach for HRV estimation in BOLD-fMRI studies. Statistical tests confirmed significant differences between configurations.

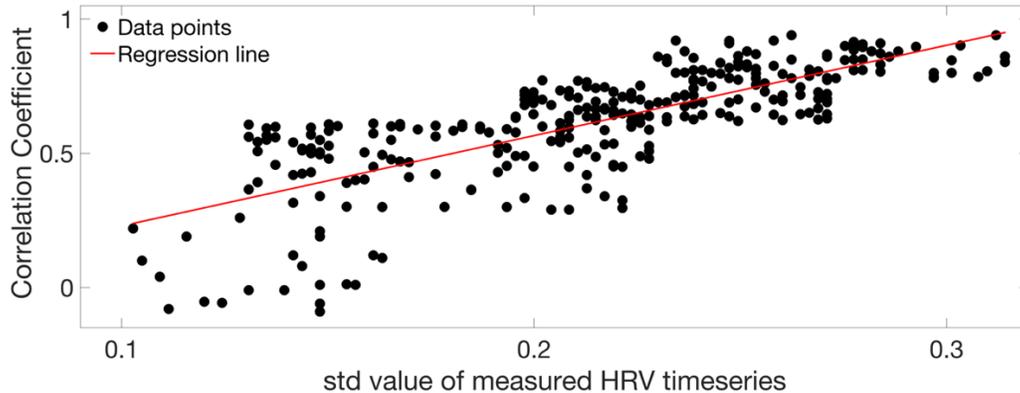

**Figure 4.** Quantitative analysis of HRV time-series variability and model reconstruction accuracy: This graph quantifies the correlation between the standard deviation of the measured HRV time-series and the Pearson correlation coefficients with the reconstructed HRV. The positive correlation demonstrates that higher variability in the HRV time-series is associated with improved accuracy in their reconstruction by the model. This trend underscores the model's capacity to effectively capture dynamic physiological signals, particularly those with greater amplitude fluctuations.

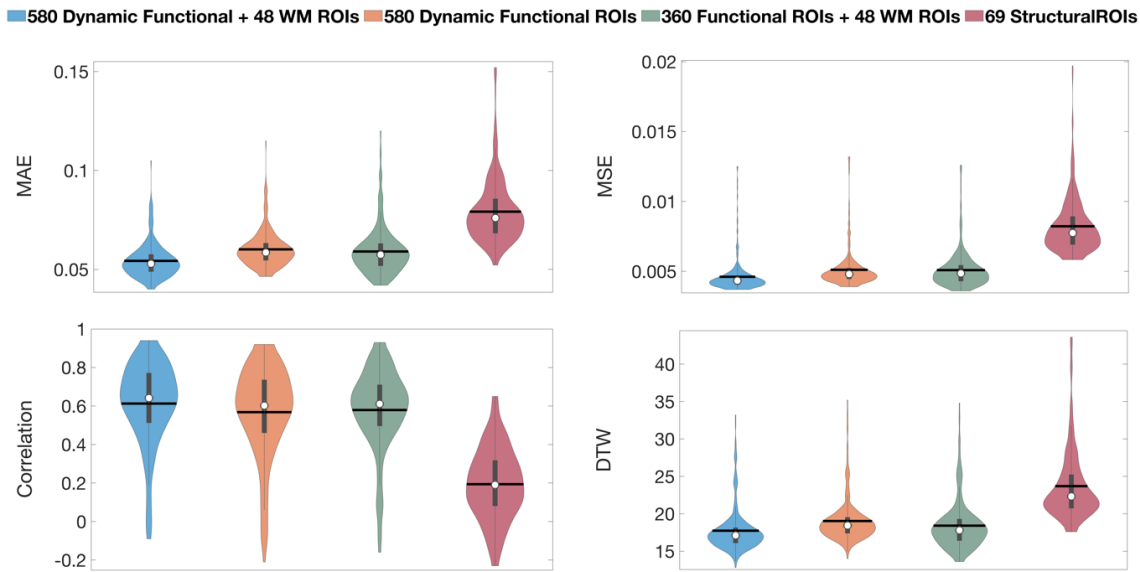

**Figure 5.** Comparative Performance of Different HRV Input Configurations. This figure presents violin plots depicting the performance of the machine learning model in reconstructing HRV across various input configurations. The configuration combining Dynamic Functional ROIs with White Matter demonstrates the highest effectiveness, underscoring the importance of integrating dynamic functional regions and white matter for HRV estimation from BOLD-fMRI. In the graphs, the white circle represents the median, and the black line indicates the mean.

**Discussion**

This study introduces an innovative method for estimating HRV from BOLD-fMRI data using a hybrid model of 1D-CNN and GRU. Our findings indicate that the dFC-based brain atlases significantly outperform static or anatomical ROIs by effectively capturing the complex temporal fluctuations of HRV.

Incorporating white matter regions substantially enhanced HRV accuracy, supporting previous studies that suggest the inclusion of non-cortical areas improves physiological confound correction [10]. This integration aids in detecting global physiological fluctuations, enhancing HRV estimation accuracy across both gray and white matter.

We hypothesized that combining dynamic ROIs, white matter, and a hybrid network would improve HRV accuracy by at least 5%, and this was confirmed with over 8% improvement in correlation and other performance metrics such as MAE and MSE. Despite these advances, challenges remain, particularly with stable HRV pattern detection due to the limitations of the sliding window technique and input variability.

Future efforts will focus on refining the network architecture and improving ROI selection methods. Testing the model on larger and more diverse datasets could also provide further insights and enhance the model's generalizability. These enhancements are critical as they highlight the potential of machine learning to transform neuroimaging analysis, setting new directions for physiological signal estimation from BOLD-fMRI data.